# Title: Ultra-long-living magnons in the quantum limit

**Authors:** Rostyslav O. Serha[1,2]*, Kaitlin H. McAllister[1,2,3], Fabian Majcen[1,2], Sebastian Knauer[1], Timmy Reimann[4], Carsten Dubs[4], Gennadii A. Melkov[5], Alexander A. Serga[6], Vasyl S. Tyberkevych[7], Andrii V. Chumak[1]**, Dmytro A. Bozhko[3]***

**Affiliations:**

[1]Faculty of Physics, University of Vienna; Vienna, 1090, Austria.

[2]Vienna Doctoral School in Physics, University of Vienna; Vienna, 1090, Austria.

[3]Center for Magnetism and Magnetic Nanostructures, Department of Physics and Energy Science, University of Colorado Colorado Springs; Colorado Springs, CO 80918, USA

[4]INNOVENT e.V. Technologieentwicklung; Jena, 07745, Germany.

[5]Faculty of Radiophysics, Electronics, and Computer Systems, Taras Shevchenko National University of Kyiv; Kyiv, 01601, Ukraine

[6]Fachbereich Physik and Landesforschungszentrum OPTIMAS, Rheinland-Pfälzische Technische Universität Kaiserslautern-Landau; Kaiserslautern, 67663, Germany.

[7]Department of Physics, Oakland University; Rochester, MI 48309, USA

Corresponding authors.
*Email: rostyslav.serha@univie.ac.at

**Email: andrii.chumak@univie.ac.at

***Email: dbozhko@uccs.edu

**Abstract:** Solid-state platforms based on bosonic quasiparticles offer a compelling route toward on-chip quantum information technologies scalable to nanometer dimensions. Coherence time, a key figure of merit for any quantum system, is fundamentally limited by the lifetime of quasiparticles that store quantum information. For magnons—bosonic excitations of collective magnetization dynamics—it has long been reported that their lifetime does not exceed a few hundred nanoseconds, placing a stringent constraint on their use in quantum architectures. Here, we demonstrate magnon lifetimes exceeding 18 µs. Experiments performed on single-crystal yttrium iron garnet spheres cooled to 30 mK reveal relaxation times of short-wavelength magnons nearly two orders of magnitude longer than previously observed. These findings overturn the established view of magnon dissipation limits, positioning magnons as viable, long-lived information carriers for solid-state quantum computing.

**Introduction:**

Quantum information processing relies fundamentally on quantum coherence, a property that quantifies the preservation of quantum states over time, enabling computational operations, data storage, and transmission in quantum systems (*1*, *2*). Among various physical systems proposed for quantum information applications, solid-state platforms offer nanometer scalability and integrability advantages but face intrinsic challenges, particularly short coherence times associated with environmental interactions and quasiparticle decay (*3–6*). In such platforms, a variety of bosonic quasiparticles—including photons (*7*), phonons (*8*), polaritons (*9*), and magnons (*10–14*) (quasiparticles representing collective spin excitations)—have been explored as potential carriers of quantum information, each offering distinct coherence properties and coupling mechanisms.

Magnons, particularly at gigahertz frequencies, are one of the most promising boson quasiparticles for quantum information carriers due to their inherent wave-like quantum behavior, tunability, rich nonlinear and nonreciprocal physical phenomena (*15–28*), and integrability into hybrid quantum systems (*29–36*). Theory further positions magnons as quantum interconnects and probes of solid-state quantum effects, predicting steady-state squeezing, photon–magnon entanglement, and nonclassical state preparation in cavity-hybrid and magneto-mechanical architectures.

Recent advances in hybrid quantum systems have demonstrated the generation and detection of single magnons using superconducting transmon qubits (*33*, *34*, *37*). This progress, along with the development of long-distance magnon transport (*38–41*) in high-quality yttrium iron garnet ($Y_3Fe_5O_{12}$, YIG, (*42*, *43*)), opens the possibility of spatially separating single-magnon sources and detectors. Such separation could enable magnon-mediated quantum gates capable of entangling distant qubits—an advantage over existing systems, where entanglement is typically limited to nearest neighbors. However, the feasibility of such gates depends critically on magnon lifetimes. For example, with qubit-magnon coupling times on the order of 200 ns (*33*), a magnon lifetime of a hundred (*37*) or a few hundred nanoseconds (*12*, *39*, *44–48*) would permit coherent interaction with only two qubits. Thus, the sub-microsecond lifetimes typically observed to date have remained a major bottleneck for scalable magnon-based quantum technologies (*12*, *46*, *47*, *49*, *50*). Notably, even such short lifetimes can sustain persistent magnetic coherence exceeding 10 μs (*48*) and magnon propagation lengths beyond 100 μm (*41*).

In this study, we report a breakthrough observation of magnons exhibiting unprecedentedly long lifetimes—from 5 μs up to 18 μs—depending on the purity of the single-crystal YIG spheres used in the measurements. The magnons were parametrically excited at gigahertz frequencies and at short wavelengths on the order of one micrometer. The magnon lifetime was determined by measuring the threshold of the nonlinear downward conversion of the externally driven spatially uniform mode of magnetization precession. Contrary to the theoretical prediction for an ideal crystal, the magnon lifetime does not increase to infinity but saturates at temperatures below about 100 mK at a level dependent on the purity of YIG, suggesting the possibility of achieving even longer lifetimes for higher purity materials.

**Results:**

Magnetic samples host a rich spectrum of magnon modes, among which the uniform precession, or Kittel's mode, is commonly employed due to its ease of excitation and characterization by Ferromagnetic Resonance (FMR) technique (*11*, *51*) (see Fig. 1A). However, the lifetime of the uniform mode is significantly influenced by the quality of the sample's surface and is generally much shorter than that of short-wavelength dipole-exchange magnons (DEMs). For this reason, DEMs are of particular interest for advanced applications, especially in quantum technologies, where extended magnon coherence times are crucial. In this study, we focus on exploring and utilizing these highly coherent DEMs. While we employ the standard FMR technique for the initial characterization of our samples, we directly measure the DEM lifetime by determining the threshold of parametric three-magnon decay (see Fig. 1B).

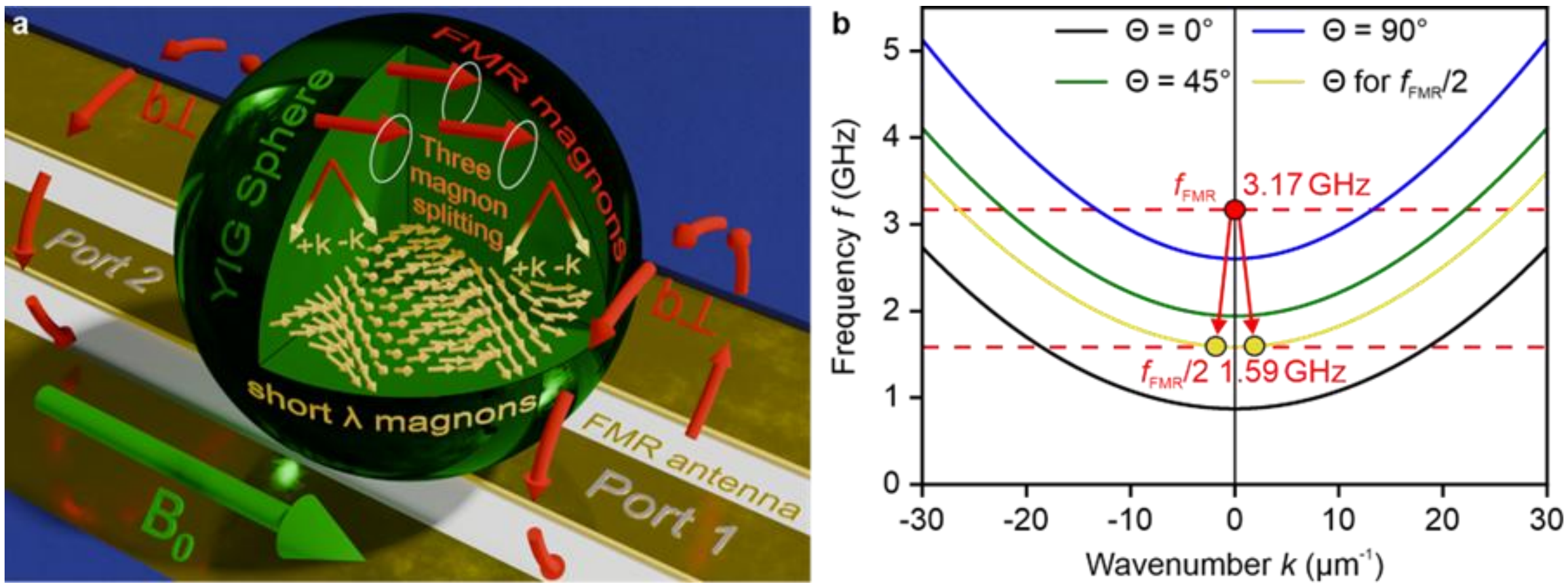

**Fig. 1. Geometry of the experiment.** (**A**) Yttrium Iron Garnet spheres of 0.3 mm diameter are placed on top of a coplanar waveguide. Magnetic field $B_0$ is applied along the conductor, ensuring a transverse direction of the microwave excitation field $b_\perp$, resulting in the most efficient excitation of the uniform precession mode (FMR). A glass spacer has been introduced to minimize the influence of coupling on the measured linewidth. The experiments were carried out in a broad range of temperatures from 300 K to 30 mK. The full detailed description of the experimental arrangement is presented in the Supplementary Materials. (**B**) Schematic diagram of the 3-magnon splitting process. The initial $k = 0$ magnon of uniform precession (FMR at 3.17 GHz) splits into two secondary magnons at half the FMR frequency of 1.59 GHz. The angular momentum conservation and symmetry of the magnon spectrum yield equal and opposite wavevectors of secondary magnons. The wavevectors of the secondary magnons are located around 3 rad/μm and lie within the DEM spectral region.

The YIG samples for our study were chosen to be spheres (because of the minimal surface-to-volume ratio) of a diameter of 300 micrometers, as in the pioneering quantum magnonics experiments (*33*, *37*). Moreover, such samples are free of the set of drawbacks associated with substrates used for the growth of YIG thin films (*52*, *53*). To study the magnon damping behavior at millikelvin temperatures and determine the role of impurities, we investigated three YIG samples grown using materials of different purity, ranging from common mass-product quality (Sphere 1) to a purer quality (Sphere 2) and ultra-pure quality (Sphere 3).

Both preliminary characterization and DEMs lifetime measurements utilized an arrangement presented in Fig. 1A. To obtain reference data for the damping of the Kittel mode, the measurements were performed using the FMR technique with sufficiently low applied powers to keep magnon excitation in the purely linear regime. The obtained FMR linewidths $\Delta B$ (see Fig. 2A), which are inversely proportional to the magnon lifetimes ($\tau = 1/\gamma\Delta B$), agree well with the previously reported behavior as a function of temperature (*45*, *46*). It is known that for long wavelength excitations, and especially for the Kittel mode, one of the main contributors to the linewidth is the sample's surface quality (*54*), as surface defects result in centers of two-magnon scattering. For samples 1 and 2, we measured approximately the same linewidth of about 0.08 mT (@ 3.17 GHz, $\tau = 70$ ns) at room temperature, which suggests a similar surface finish. For the ultrapure sample 3, the linewidth is about twice as small. All samples except the ultra-pure one exhibit a pronounced peak in linewidth around 50 K, which is attributed to the magnon relaxation on rare-earth ion impurities(*12*, *47*, *55*). Further details about samples, FMR, and other measurement techniques used in this study can be found in the Supplementary Materials.

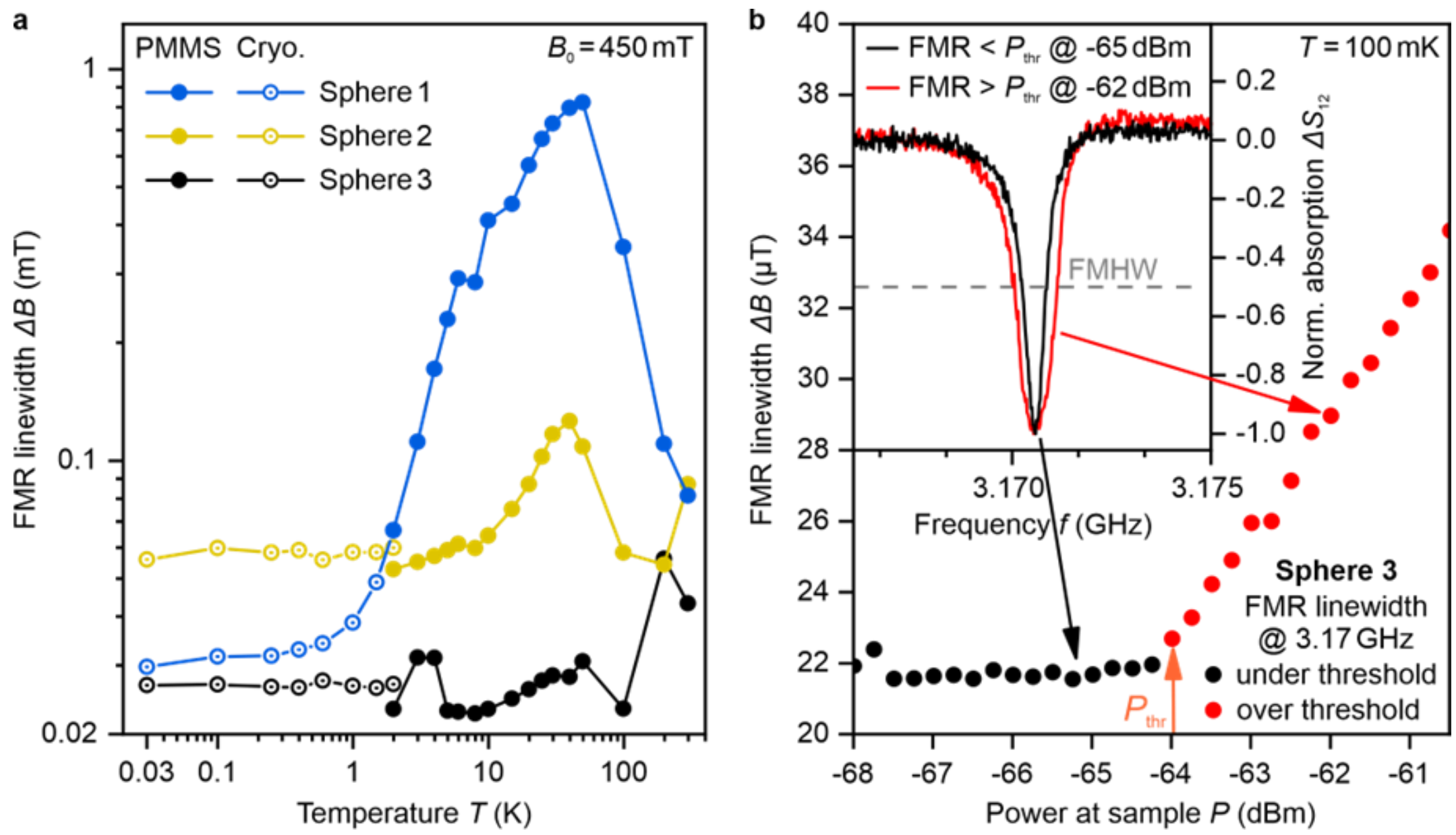

**Fig. 2. Ferromagnetic resonance measurements and 3-magnon splitting (down-conversion) process.** (**A**) FMR linewidth as a function of temperature for three different YIG spheres on double logarithmic axes, measured in the linear regime. The solid dots represent measurements performed in the Physical Property Measurements System (PPMS), while the hollow dots correspond to measurements taken in the dilution refrigerator. (**B**) Exemplary FMR lines measured at 3.17 GHz for two different applied microwave power levels and the dependence of the FMR linewidth on the applied power. A clear threshold-like increase in the linewidth indicates an onset of the 3-magnon instability.

To measure the lifetime of DEMs we employed the nonlinear process of parametric downconversion, also known as three-magnon decay (see Fig. 1B). This process involves the splitting of a single magnon at the FMR frequency (3.17 GHz in Fig. 1B) into a pair of DEMs, each at half that frequency (1.59 GHz in Fig. 1B). This phenomenon is permitted by the energy and momentum conservation laws only when half of the FMR frequency falls within the continuum of DEM modes. For spherical samples, this condition is met at low magnetic fields $H_0 < 2M_0/3$. When the rate of three-magnon decay surpasses the DEM damping rate, the population of DEMs grows exponentially. This exponential growth leads to an additional energy transfer to the DEMs, which subsequently causes an increase in the linewidth of the Kittel mode. Consequently, precise measurements of the threshold power for three-magnon decay offer a direct and accurate method for determining the lifetime of short-wavelength DEMs.

Example measurements of the linewidth of Kittel's mode in the nonlinear regime are shown in Fig. 2B. The linewidth stays constant up to the threshold value $P_{\mathrm{thr}} = -64$ dBm, and linearly increases in the super-threshold region due to energy transfer to parametric DEMs. The temperature dependence of the threshold power $P_{\mathrm{thr}}$ for all three studied spheres is shown in Fig. 3A. A pronounced reduction of about 30 dB is observed as the temperature decreases from room temperature down to 30 mK. This dependence is a direct manifestation of the decrease in the relaxation parameter of the secondary magnons. The absolute value of the threshold power at mK temperatures is different for different spheres, which is also an indication of the different magnon lifetimes, as well as evidence of a strong influence of the extrinsic relaxation channel associated with impurities on the magnon's lifetime, as we will discuss below.

The threshold power $P_{\text{thr}}$ can be directly related to the lifetime $\tau$ of parametrically excited DEMs (see Supplementary Information for details):

$$\tau = \frac{6\sqrt{V\Gamma_0\omega_0^3}}{\sqrt{P_{\text{thr}}A\omega_M\mu_0}\gamma(2\omega_M - 9\omega_0)\sin(2\theta_k)}. \quad (1)$$

Here $V$ is the sample volume, $\Gamma_0 = \gamma\Delta B$ is the linear FMR linewidth, $\omega_0$ is the FMR frequency, $A$ is the FMR absorption ratio, $\omega_M = \gamma\mu_0 M_0$, $\gamma$ is the modulus of the gyromagnetic ratio, $\mu_0$ is the vacuum permeability, and $\theta_k$ is the polar angle of the excited DEM determined by the magnon dispersion relation and energy conservation law.

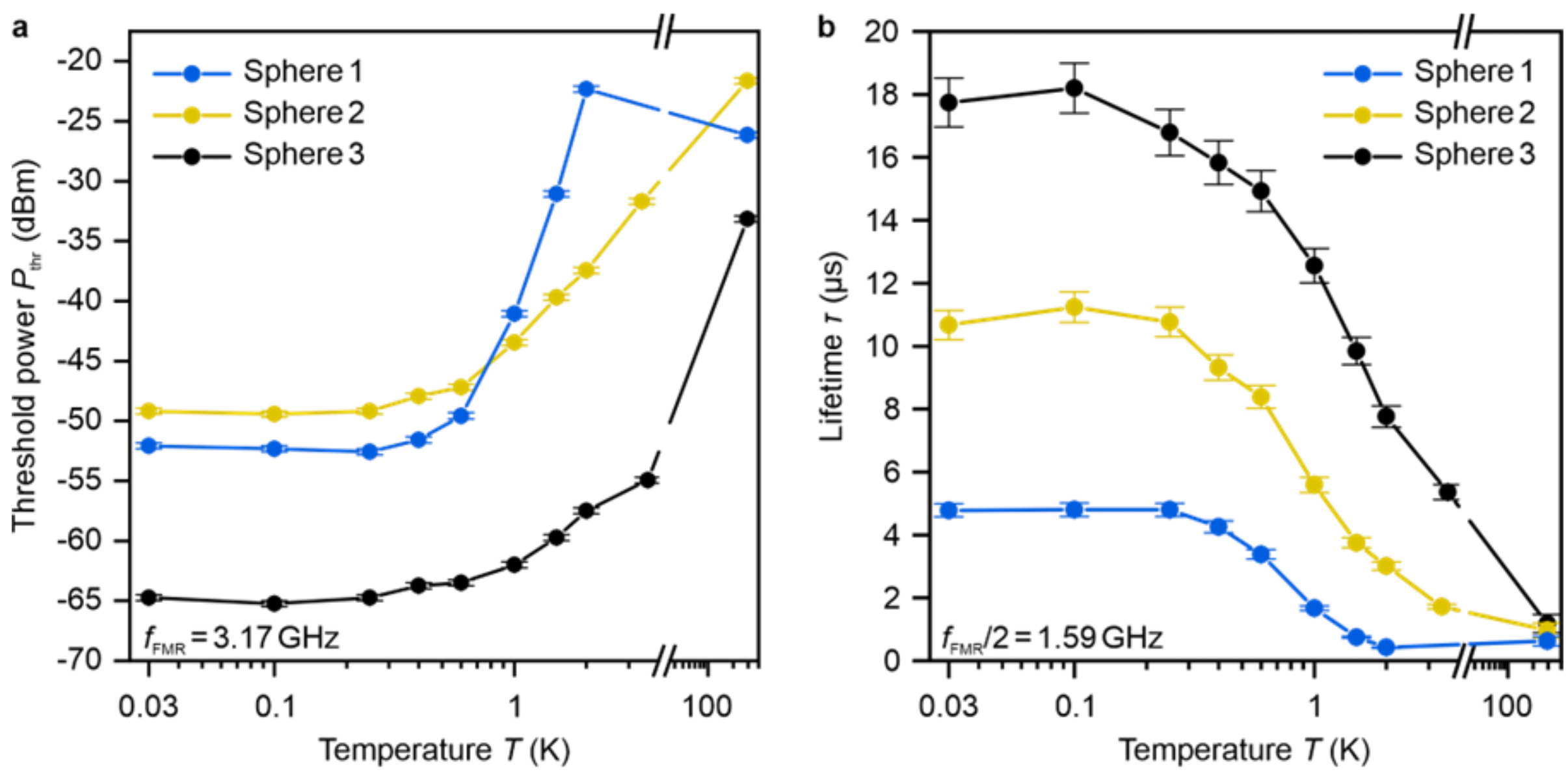

**Fig. 3. Threshold RF power and lifetime of secondary magnons vs temperature.** (**A**) Threshold rf power $P_{\text{thr}}$ as a function of temperature on a logarithmic x-axis for three different YIG spheres at the FMR frequency of 3.17 GHz. (**B**) Lifetime $\tau$ of dipole-exchange magnons with the frequency half of the FMR frequency vs the temperature $T$ on a logarithmic x-axis for three different YIG spheres. The procedures used for the determination of the DEM lifetime $\tau$ are described in the Supplementary Materials.

Using this expression, we can transform data from Fig. 3A into the magnon lifetime shown in Fig. 3B. We would like to note that the room temperature datapoint matches well the lifetimes obtained by analyzing FMR data of current and previous works—the values are of the order of 1 μs. This also indicates that at room temperature, we are dealing with strong contributions from intrinsic multi-magnon relaxation processes. Below 4 K, the magnon lifetime starts to grow. Even for the least pure Sphere 1, we measure a magnon lifetime of 4.5 μs at 30 mK, which significantly exceeds all previously reported values. For Sphere 2, we determined the magnon lifetime to be 11 μs, and for the ultra-pure Sphere 3, the measurements resulted in a whopping 18 μs. In terms of the linewidth of the secondary magnons $\Delta B_k$, this lifetime corresponds to the value of 0.63 μT. Lifetime dependencies saturate and do not increase further below temperatures of about 100 mK.

**Discussion:**

The observed here magnon lifetimes of up to 18 μs are exceeding all previously reported values at any temperature by about two orders of magnitude. This breakthrough is a result of several underlying factors discussed below. The first fundamental factor is that here we excite short-wavelength dipole–exchange magnons. These magnons appear to be insensitive to surface defects, allowing them to bypass the two-magnon scattering that limits long-wavelength and FMR magnon damping both in classical as well as quantum-limited experiments (*12*, *45*, *46*, *56*, *57*). The second factor is that intrinsic damping pathways, primarily associated with the magnon-phonon and magnon-magnon scattering, become inhibited in the millikelvin regime. And indeed, as it is evident from Fig. 3B, the damping decreases monotonically with decreasing temperature for all samples due to the decrease in thermal quasiparticles population down to temperatures of approximately 100 mK. This temperature can be associated with the thermal energy of $k_B T \approx 2$ GHz, which is close to the experimentally measured frequency of 1.59 GHz, and, therefore, the magnon and phonon bath populations below this temperature are strongly suppressed, diminishing both scattering channels (further peculiarities of the intrinsic temperature-dependent relaxation behavior can be found in the Supplementary Information).

As other decay channels are suppressed, the remaining decay—responsible for the saturation of the lifetime below 100 mK in Fig. 3B for samples of different purity—originates from rare paramagnetic centers in the crystal lattice, i.e. fluctuating local moments (*58*–*60*). While the absolute concentration of these impurities can be made very small, even a few defects can dominate the relaxation landscape once intrinsic processes are eliminated. Measurements across our set of spheres show that fluctuators' influence weakens as the crystal quality improves, leading to long relaxation times saturating at low temperatures. Remarkably, that even with this saturation, the lifetime of dipole–exchange magnons below 100 mK reaches previously unattained values—about 5 μs even in the least pure sphere. As the fluctuation rates of these impurities decrease as the temperature approaches zero (*58*), one expects that, in the absence of extrinsic decoherence channels, the lifetime of dipole–exchange modes would not exhibit saturation at all but would instead continue to grow. However, in our experiments we have not achieved temperatures below 30 mK, at which this damping mechanism could be substantially suppressed (*61*).

These observations highlight a crucial point: the coherence of short-wavelength magnons is not fundamentally constrained by the mechanisms that limit the uniform mode lifetime. Instead, it is governed by a hierarchy of processes whose relative importance changes dramatically as temperature is reduced. Our results, therefore, establish a clear materials-science route for extending magnon coherence further—namely, through continued reduction of impurity concentrations and improved control over crystal quality.

The 18 μs lifetime reported here places magnon coherence on par with that of typical transmon superconducting qubits in quantum processor networks (*62*), opening pathways for practical quantum technologies where short-wavelength magnons function as long-lived quantum information carriers. This level of coherence reshapes the role of magnons in hybrid quantum architectures—from lossy intermediaries to robust quantum memories and low-loss waveguide links capable of mediating non-local interactions across chip-scale distances. It establishes magnons as one of the most perspective quasiparticles for modern quantum solid-state physics. When resonantly coupled to superconducting circuits (*31*, *33*, *34*, *56*, *63*, *64*), such long-living magnons can mediate non-local entanglement among hundreds of qubits along a common waveguide (*65*) and act as a programmable on-chip quantum bus for entangling gates (*66*). In doing so, they supply a missing hardware building block for modern superconducting processors and advance the challenging route toward scalable quantum computing (*67*, *68*).

## Materials and Methods

Dilution refrigerator

Our experimental setup is based on a cryogen-free dilution refrigeration system (Bluefors LD250) capable of reaching base temperatures below 10 mK at the mixing chamber stage. The sample space maintains a base temperature of about 20 mK, although it can be heated up to about 30 mK during operation by changing the external magnetic field. At these temperatures, thermal excitations of gigahertz frequency magnons and phonons are still sufficiently suppressed. The setup allows measurements at temperatures from 30 mK to 3.6 K. The external magnetic field is applied via a superconducting solenoid, providing a homogeneous magnetic field around the sample.

For cryogenic measurements in the dilution refrigerator, the transmitted signals are acquired using a high-performance vector network analyzer (VNA) (Anritsu VectorStar MS4647B). The input signal is highly attenuated (40 dB to 60 dB) before being transmitted to the sample through Port 1 and collected from Port 2 after amplification by a room-temperature amplifier (LNF-LNR4_14C_SV). The microwave lines inside the cryostat consist of a combination of high-frequency copper and superconducting wiring, identical for both input and output paths.

PPMS

To verify the purity and determine the magnetic damping across a wide temperature and frequency range, we used a commercial Physical Property Measurement System (Quantum Design PPMS) in combination with the VNA (Rhode & Schwarz ZVA40) to perform FMR measurements on our three samples. This setup enables FMR measurements from 300 K down to 2 K, magnetic fields of up to 9 T and frequencies up to 40 GHz within the same experimental configuration.

Room temperature setup

For the room-temperature (RT) setup, the VNA (Rhode & Schwarz ZNA67) is connected to an H-frame electromagnet (GMW 3473-70) with an adjustable air gap and is powered by a bipolar power supply (BPS-85-70EC, IECO). The magnet poles have a diameter of 10 cm with an interpole distance of 4 cm, providing a sufficiently uniform biasing magnetic field across the sample. For room-temperature measurements, no external amplifier was necessary. The microwave lines to and from the sample were symmetrical, ensuring accurate knowledge of the power delivered to the sample.

Microwave excitation circuit

In all setups, we use a U-shaped grounded coplanar waveguide (CPW) ferromagnetic resonance (FMR) antenna (NanOsc Instruments CPW PPMS IP) to excite and measure the FMR absorption in the magnetic system of the sample. The signal width is 200 μm with 150 μm spacing on either side at the sample position. The thickness of the dielectric layer is 200 μm with a dielectric constant of 3.26. The vias are 200 μm wide with a center spacing of 400 μm.

Samples

The samples used in this experiment consist of three 300 μm-diameter single-crystal yttrium iron garnet (YIG) spheres, each originating from different sources and exhibiting varying levels of purity. The first sample is a commercially available YIG sphere (Micro Spheres) with the lowest degree of purity with paramagnetic rare earth impurities from > 1.3 ppm (part per million), named Sphere 1. The second sample, provided by INNOVENT e.V. Technologieentwicklung Jena, Germany, was prepared from a single crystal grown in a high temperature solutions applying the slow cooling method. The purity of the yttrium oxide ($Y_2O_3$) used was 99.9999%, based on the total content of rare earth oxides, in order to minimize the incorporation of rare earth relaxer elements into the garnet lattice. As a result, the high-purity YIG sphere, named Sphere 2, contains only around 1.3 ppm of paramagnetic rare earth elements. The third sample, named Sphere 3, is a YIG sphere ground from a single crystal of the highest purity (with significantly less paramagnetic rare earth elements than 1.3 ppm, sourced from the collection of the Taras Shevchenko National University of Kyiv, Ukraine, and was shaped into a sphere at INNOVENT e.V. Technologieentwicklung Jena. The spheres were made from cubes - cut from the crystals - by grinding individual samples to diameters of (300 ± 10) μm. The asphericity was better than one percent. Finally, an additional polishing step allows the preparation of a smooth surface without scratches and holes.

The magnetic damping properties of the three YIG spheres were characterized by measuring the FMR linewidth over a frequency range from 3 GHz to 20 GHz and performing linear Gilbert fits, as shown in Fig. S1. The extracted Gilbert fit parameters — the Gilbert damping α and the inhomogeneous linewidth broadening $\Delta B_0$ — are summarized in Table S1. All samples were oriented with <111> axis along the biasing magnetic field. This magnetization direction was chosen as the damping associated with impurities is minimal in such an orientation (*69*).

At 300 K, Sphere 3 exhibits the narrowest linewidths, indicating the best combination of both Gilbert fit parameters. This result confirms that Sphere 3 is of the highest quality among the three spheres (see Fig. 1A). While Sphere 2 has the smallest Gilbert damping parameter, it also shows the largest inhomogeneous linewidth broadening $\Delta B_0$, making its overall linewidths comparable to Sphere 1 at room temperature.

At the lowest measured temperature of 30 mK, the FMR linewidths of Spheres 1 and 3 decrease significantly, particularly at higher frequencies, resulting in a substantial reduction in the Gilbert damping α, as shown in Table S1. Sphere 1 improves its damping performance at low temperatures and becomes comparable to Sphere 3 in terms of both damping and linewidth broadening. In contrast, Sphere 2 shows an even lower Gilbert damping parameter at 30 mK, but its relatively large inhomogeneous linewidth broadening $\Delta B_0$ of 59.46 μT results in the largest overall linewidths among the three spheres (see Fig. 1b). In summary, Sphere 3 exhibits the lowest magnetic damping at 300 K, and this damping further decreases at 30 mK to a Gilbert damping $\alpha = 2.05 \cdot 10^{-5}$ and inhomogeneous linewidth broadening $\Delta B_0 = 7.61$ μT, confirming its superior quality across the temperature range.

The purity order of the YIG spheres can be verified by examining Fig. 2A, which shows a comparison of the FMR linewidths for the three spheres measured at an external magnetic field of $B_0 = 450$ mT across a temperature range from 30 mK to 300 K, plotted on a logarithmic x-axis. Sphere 1 (blue) exhibits a significant increase in linewidth, peaking at around 50 K before gradually freezing out as the temperature approaches the millikelvin range. This distinctive behavior indicates a substantial level of rare-earth impurities in the YIG crystal, confirming that this sphere has the lowest purity among the samples investigated.

Despite this, it is notable that the total linewidth of Sphere 1 at millikelvin temperatures, when the damping increase due to rare-earth impurities is frozen out, is smaller than that of Sphere 2. This suggests that, despite the impurity content, Sphere 1 is a high-quality single crystal. Sphere 2 (yellow) displays a similar trend but with a much smaller increase in linewidth at comparable temperatures. This indicates a lower concentration of rare-earth impurities compared to Sphere 1. Sphere 3 (black) demonstrates remarkable behavior, with only a barely visible increase in linewidth across the entire temperature range. Moreover, it exhibits the lowest linewidth at all measured temperatures, making it the purest YIG sample in this investigation.

FMR and parametric threshold determination

To ensure that the FMR absorption remained in the highly undercoupled regime to the external microwave line, glass spacers with a thickness between 350 μm and 450 μm were placed between the stripline antenna and the YIG sphere. This adjustment was necessary to optimize the coupling and prevent any broadening of the linewidth due to external losses.

Measurements were conducted at cryogenic temperatures for two FMR frequencies, 3.17 GHz and 3.87 GHz. At room temperature (RT), however, three-magnon scattering could only be observed at 3.17 GHz due to the lower saturation magnetization of YIG at higher temperatures. This highlights the temperature-dependent nature of the scattering process and its sensitivity to the material properties under different conditions.

The three YIG spheres were mounted on the edge of a 200 μm-thick glass spacer, which served as a sample holder, while an external magnetic field was applied. This ensured that the crystallographic orientation of the spheres could be estimated as ⟨111⟩ along the external field during the mounting process. The samples were then placed on the FMR antenna in such a way that the applied magnetic field remained parallel to both the spheres' easy magnetization axis and the FMR stripline

The FMR absorption spectrum was measured by applying a radio-frequency (RF) sweep through a stripline using a VNA, which aqquired the transmitted signal between Port 2 and Port 1 as the scattering parameter $S_{12}$. To isolate the FMR response at a specific magnetic field, the transmission $S_{12}$ was measured both at the target field and at a reference field offset by approximately 1-5 mT. The difference between these measurements, $\Delta S_{12}$, provides a clear FMR spectrum with nonmagnetic background contributions effectively suppressed. The resonance frequency $f_{\mathrm{FMR}}$ and full width at half maximum (FWHM) $\Delta B$ were extracted by fitting the resonance profile with a Lorentzian model.

Determining the power thresholds for three-magnon scattering was essential for calculating the magnon lifetimes presented in the main article. To achieve this, the FMR spectrum at each temperature was measured with small power increments, using a step size of 0.25 dBm. Reference measurements were taken with an external magnetic field step of 1 mT to minimize the influence of field variations on the measured spectrum and to reduce the overall measurement time. Each sweep consisted of 6001 points with an intermediate frequency bandwidth (IFBW) of 3 kHz and a frequency window of 140 MHz, resulting in a frequency step size of 23 kHz. This resolution was necessary to properly capture the narrow FMR peaks of the YIG spheres. Additionally, various IFBW settings, ranging from 30 Hz to 30 kHz, were tested to ensure that the sweep speed did not affect the power threshold measurements.

To accurately determine the power reaching the sample, a calibration measurement was performed to account for power loss through the symmetrical microwave lines. The total power at the sample was calculated by subtracting half of the measured power loss at the corresponding frequency. This calibration ensured that the power thresholds were correctly referenced to the actual power delivered to the sample, improving the reliability of the extracted parameters.

Figure 2B illustrates the process used to identify the power threshold for the onset of three-magnon scattering, using an example measurement at 100 mK with an FMR frequency of 3.17 GHz. The FMR peaks were fitted with Lorentzian functions to extract both the absorbed power and the linewidth, the latter of which was then plotted as a function of the power applied to the sample. In the linear regime, the linewidth and peak shape remained constant as the input power increased, as shown by the black data points in Fig. 2B.

However, once the power threshold was reached, marked by the orange arrow in Fig. 2B, a clear transition to the nonlinear regime occurred. At this point, the linewidth broadened significantly, the peak amplitude decreased, and the FMR signal became distorted, eventually losing its Lorentzian profile at higher power levels. This transition indicates the onset of parametric instability, where the system moves beyond stable magnon excitation into the three-magnon scattering regime. The nonlinear regime is depicted by the red data points in Fig. 2B.

To further illustrate this transition, the inset of Fig. 2B shows normalized FMR spectra below (black) and above (red) the power threshold. The comparison highlights the broadening of the FMR peak once the threshold is surpassed. The power level at which this nonlinear broadening first appears is identified as the threshold point.

The constant behavior of the FMR linewidth $\Delta B$ in the under-threshold regime and its linear increase in the over-threshold regime, as shown in Fig. 2B, suggests that the power threshold could be precisely determined by the intersection of two linear fits. However, in many measurements, the FMR linewidth in the over-threshold regime does not follow a general linear trend and also can exhibit significant scattering, making a precise linear fit unreliable. To maintain consistency in the evaluation, the power threshold was defined as the first data point that deviates from the constant under-threshold behavior, with an uncertainty of ±0.25 dBm. While this approach results in larger error bars for the extracted secondary magnon lifetimes, it ensures that the power threshold and corresponding lifetime values are not overestimated.

The parametric power thresholds $P_{\mathrm{thr}}$ are shown in Fig. 3A and Fig. S3A as a function of temperature for the two measured FMR frequencies 3.17 GHz and 3.87 GHz. The linewidth and the percentage of power absorption from the final FMR spectrum within the linear regime were used to obtain the critical threshold RF field $b_{\mathrm{thr}}$ (see Fig. S2A and Fig. S3A) for the calculation of the secondary magnons linewidth $\Delta B_k$ and lifetime τ, in detail presented below. The secondary magnons' linewidth $\Delta B_k$ shown in Fig. S2B and Fig. S3C.

Analytical expression for three-magnon decay threshold

The threshold value of the RF magnetic field of parametric instability (*55*) in our particular case can be written as:

$$b_{\mathrm{thr}} = \min\left\{\frac{2\omega_0\Gamma_k\Gamma_0}{\gamma\omega_{\mathrm{M}}\sin(2\theta_k)\,(3\omega_0/2+\eta k^2)}\right\}, \tag{2}$$

where $\Gamma_k = 1/\tau$ is the relaxation frequency of the parametrically excited DEMs with wavevector $k$ ($\tau$ is magnon lifetime), $\Gamma_0$ is the relaxation frequency of the FMR mode, $\omega_M = \gamma M_0$, $\theta_k$ is the angle of the DEM's wavevector with respect to the applied field direction, and $\eta$ is the exchange constant. Minimization is performed over all magnon wavevectors $k$ and angles $\theta_k$ allowed by the energy conservation law. Assuming a small dependence of the relaxation rate $\Gamma_k$ on the wavevector $k$ in the spectral region of interest, in our case the minimization results in excitation of DEMs with $k \approx 3\ \mathrm{rad/\mu m}$, for which the exchange term $\eta k^2$ is small and can be ignored, and largest allowed polar angle

$$\sin^2(\theta_k) = \frac{(2\omega_M - 3\omega_0)(9\omega_0 - 2\omega_M)}{12\omega_M(3\omega_0 - \omega_M)}. \quad (3)$$

The last step is to relate the threshold field to the experimental parameters. This can be done precisely by extracting $b_{\text{thr}}(P_{\text{thr}})$ dependence analytically from the $S_{21}$ data measured in the FMR experiment (see e.g., inset in Fig. 2B) in- and out-of-resonance:

$$b_{\text{thr}} = \sqrt{P_{\text{thr}} A \frac{4\mu_0\omega_{\text{r0}}}{\omega_{\text{M}}\omega_0 V}}, \quad (4)$$

where $P_{\text{thr}}$ is the threshold power, $A$ is the absorption ratio of the FMR, and $V$ is the volume of the YIG sphere. The experimentally determined dependence of the threshold rf field on temperature for all three spheres is shown in Fig. S2.

The explicit expression Eq. (1) for the magnon lifetime is obtained by solving Eq. (2) for $\tau$ and using Eq. (4) for $b_{\text{thr}}$.

**Acknowledgments:**

**Funding:**

This material is based upon work supported by the National Science Foundation under Award No. DMR-2338060 (DAB).

This research was funded in whole or in part by the Austrian Science Fund (FWF) Project No. 10.55776/I6568 (AVC).

Deutsche Forschungsgemeinschaft (DFG, German Research Foundation)—TRR 173—268565370 Spin+X (Projects B01 and B04) (AAS).

SK acknowledges the funding from the European Union's Horizon 2020 research and innovation programme under the Marie Sklodowska-Curie grant agreement No.101025758.

GAM acknowledges support from the IEEE Magnetics Society through the "Magnetism for Ukraine" initiative.

**Author contributions:**

Conceptualization: DAB, AVC

Methodology: ROS, KHM, FM, SK, TR, CD, GAM, AAS, VST, AVC, DAB

Investigation: ROS, KHM, FM, SK, GAM, AAS, VST, AVC, DAB

Visualization: ROS, KHM

Funding acquisition: AVC, DAB

Project administration: AVC, DAB

Supervision: AVC, DAB

Writing – original draft: ROS, AVC, DAB

Writing – review & editing: ROS, KHM, FM, SK, TR, CD, GAM, AAS, VST, AVC, DAB

**Competing interests:** Authors declare that they have no competing interests.

**Data and materials availability:** Data and materials used in the analysis are available in the main text or the supplementary materials. An extended datasets and materials are available from the corresponding authors upon reasonable request.

# Supplementary Materials for

## Ultra-long-living magnons in the quantum limit

Rostyslav O. Serha, Kaitlin H. McAllister, Fabian Majcen, Sebastian Knauer, Timmy Reimann, Carsten Dubs, Gennadii A. Melkov, Alexander A. Serga, Vasyl S. Tyberkevych, Andrii V. Chumak, Dmytro A. Bozhko

Corresponding authors: rostyslav.serha@univie.ac.at, andrii.chumak@univie.ac.at, dbozhko@uccs.edu

**The PDF file includes:**

Supplementary Text
Figs. S1 to S3
Table S1

**Supplementary**

Three-magnon scattering at 3.87 GHz

In addition to the secondary magnon lifetime results presented in Fig. 2 for measurements at an FMR frequency of 3.17 GHz, we provide supplementary measurements at a higher frequency of 3.87 GHz in Fig. S3 to further support the conclusions in the main text. Measurements at 3.87 GHz could not be performed at room temperature (RT) due to the lower saturation magnetization $M_s$ of YIG at higher temperatures, which makes the three-magnon scattering process inaccessible at this frequency. However, for consistency and ease of comparison, Fig. S3 maintains the same scale as Fig. 2.

Figure. S3A shows the parametric power threshold $P_{thr}$ as a function of temperature and Fig. S3B shows the corresponding threshold RF field $b_{thr}$, plotted on a logarithmic x-axis. Among the three spheres, Sphere 3 (black) exhibits the lowest threshold power and RF field across the entire temperature range, with $b_{thr}$ decreasing as the temperature drops and saturating below 600 mK. The behavior of Sphere 2 (yellow) and Sphere 1 (blue) follows a similar trend, but both show significantly larger threshold RF fields compared to Sphere 3. Notably, Sphere 1 shows a steeper decrease in $b_{thr}$ resulting in smaller threshold RF fields compared to Sphere 2 at temperatures below 1 K. This behavior can be understood by examining the FMR linewidths presented in Fig. 2A. As shown, the linewidth of Sphere 1 decreases steeply due to the freezing out of rare-earth impurities, making it smaller than that of Sphere 2 within the same temperature range.

Overall, the threshold RF fields for all spheres are larger at an FMR frequency of 3.87 GHz compared to the 3.17 GHz measurements, which is consistent with the higher magnon frequency requiring stronger driving fields for instability.

As described in the main manuscript, the lifetimes of secondary magnons are extracted from the threshold RF field values $b_{thr}$, and these are shown in Fig. S3D. The behavior of DEMs with a frequency of 1.94 GHz (corresponding to the 3.87 GHz FMR frequency) closely resembles the results for 1.59 GHz magnons presented in Fig. 2B. Sphere 1 exhibits the shortest magnon lifetimes, followed by Sphere 2, while Sphere 3 shows the longest lifetimes, reaching a maximum of approximately 14 μs. The increase in magnon lifetime saturates earlier for 1.94 GHz magnons, below 400 mK, compared to the results for 1.59 GHz magnons (around 100 mK). One possible explanation for the observed difference in saturation behavior is that, at higher frequencies, the thermal baths of magnons and phonons involved in the scattering processes described in the main article freeze out at comparatively higher temperatures.

The magnon lifetimes at 1.94 GHz are shorter than those at 1.59 GHz for all YIG spheres. Specifically, the lifetimes are reduced by approximately 12% for Sphere 1, 15% for Sphere 2 and 20% for Sphere 3. These results confirm that secondary magnon lifetimes decrease with increasing magnon frequency.

Magnon relaxation channels

Although these are the first reported measurements of $k \neq 0$ magnon lifetimes in the quantum limit, relaxation mechanisms contributing to magnon damping have been studied previously down to a Kelvin range of temperatures (*1–8*). The rather high election band gap of 2.8 eV of YIG (*9*) precludes magnon-electron relaxation processes (*10*). The dominant mechanism defining the FMR linewidth $\Delta H_0$ is elastic two-magnon scattering, which is temperature-independent. As most of the defects are associated with surface roughness, the best FMR linewidths could be achieved for samples with the highest volume-to-surface ratio – spheres. As a result, $\Delta H_0$ remains nearly constant with temperature (see Table S1) for the purest YIG Sphere 3, where magnon-ion relaxation is greatly reduced.

On the contrary, for the short-wavelength magnons, we observe a pronounced temperature dependence of the threshold power for three-magnon splitting, which suggests that their relaxation, often referred to by the linewidth $\Delta H_k$, is not dominated by two-magnon scattering(*11*) and therefore is governed by other processes. These processes include intrinsic effects of magnon-magnon (also known as spin-spin) (*12*, *13*) or magnon-phonon (or spin-lattice) (*14*, *15*) interactions, as well as extrinsic effects of coupling to magnetic moments of impurities (*16*, *17*). Three-magnon splitting is forbidden for the secondary magnons due to their location close to the bottom of the spin-wave spectrum (see Fig. 1B). Other nonlinear relaxation processes such as three-magnon confluence and four-magnon scattering are exponentially ineffective, as they require large magnon populations (either thermal or stimulated) spectrally close to the magnons in question, which is not the case for our experiments done in the $T \rightarrow 0$ limit. This statement is further supported by the fact that the damping behavior below 100 mK experiences saturation for all samples. This temperature is associated with the thermal energy of $k_B T \approx 2$ GHz, which is close to the experimentally measured frequency of 1.59 GHz, and, therefore, the magnon (as well as phonon) population below this temperature is strongly suppressed.

The direct magnon-phonon interaction processes, including the so-called Kasuya-LeCraw mechanism (*18*) available for an ideal crystal lattice and only small wavevectors, scale well with temperature as the population of the phononic bath is effectively suppressed in the $T \rightarrow 0$ limit (*15*). When magnon's group velocity is higher than the phonon's one, magnons can scatter into another magnon and a phonon – the so-called Cherenkov process. However, this effect is also suppressed for short-wavelength magnons with moderate frequencies. But despite the impression that all the damping mechanisms are becoming suppressed, the observed magnon relaxation here does not vanish at $T \rightarrow 0$, as predicted theoretically for ultra-pure YIG. In our case, the residual magnon damping can still be attributed to the intrinsic magnon interaction with the low-frequency portion of the phonon bath. Since phononic spectrum does not have a gap, this contribution will always be present at finite temperature. However, this contribution is intrinsic and therefore should be equal for all the studied spheres, which is clearly not the case in our findings – see Fig. 3B. That gives a strong hint that the main contributor to the residual damping is coming from impurities.

Despite the ultra-high purity of Sphere 3, we cannot guarantee that the residual impurity concentration was zero. Such solitary paramagnetic impurities, which include either rare-earth elements or $Pt^{4+}$, $Si^{4+}$, $Fe^{2+}$, and $Fe^{4+}$ originated from the crystal growth process (*19*), directly interact with magnons through their fluctuating magnetic moment and are called two-level-fluctuators (TLF), as proposed in Refs. (*16*, *17*). Their influence strongly depends on their own relaxation frequency, which scales with temperature and should vanish at $T = 0$ (*1*) but apparently in our case, that was not yet achieved. The design of a particular experiment to quantify TLFs' influence on short-wavelength magnons constitutes a fundamentally interesting problem, which spans outside the scope of the current work. However, our findings clearly show that 18 μs, as reported here, is not the fundamental limit of the lifetime of this type of boson and that it can be significantly increased further by optimizing the fabrication technology.

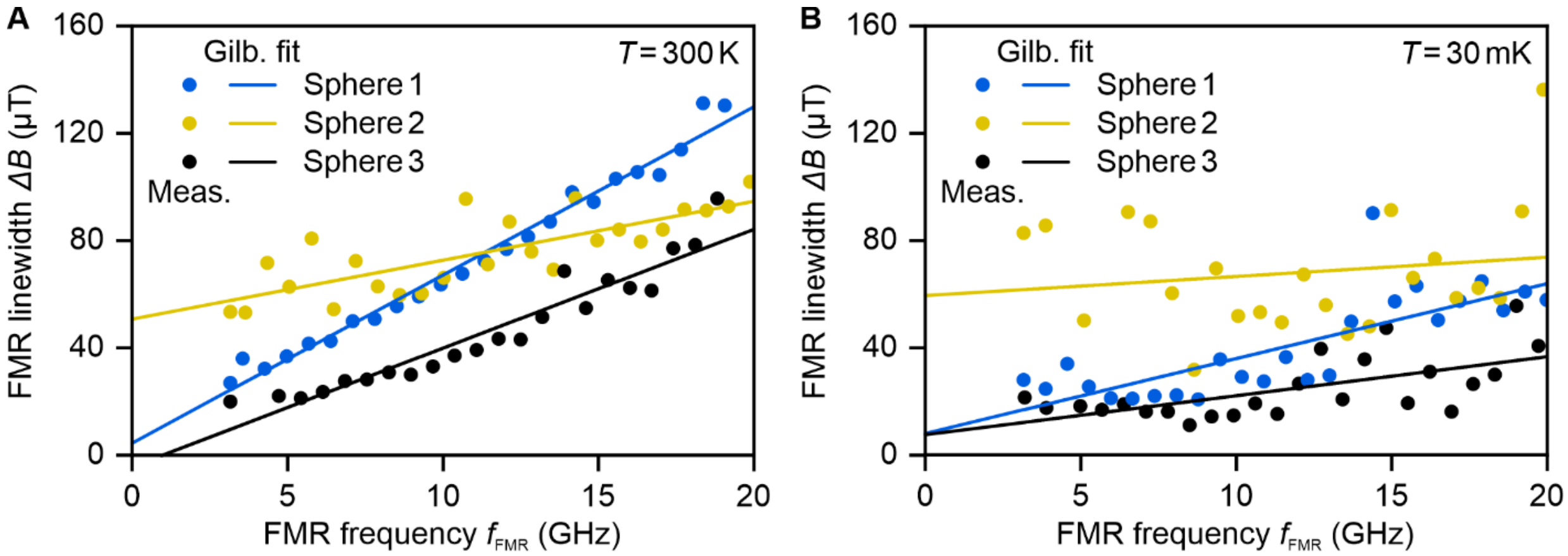

**Fig. S1.**

**FMR linewidth $\Delta B$ as a function of the FMR frequency $f_{FMR}$.** The dots represent the measured linewidths for three different YIG spheres: Sphere 1 (blue), Sphere 2 (yellow), and Sphere 3 (black), while the solid lines show the corresponding linear Gilbert fit. Panel A displays measurements taken at 300 K, and B shows measurements at 30 mK. From the measurements it becomes evident that Sphere 3 has the lowest damping parameters for both temperatures. While Sphere 2 has a smaller slope, corresponding to a lower Gilbert damping parameter α, it also has a higher $\Delta B_0$.

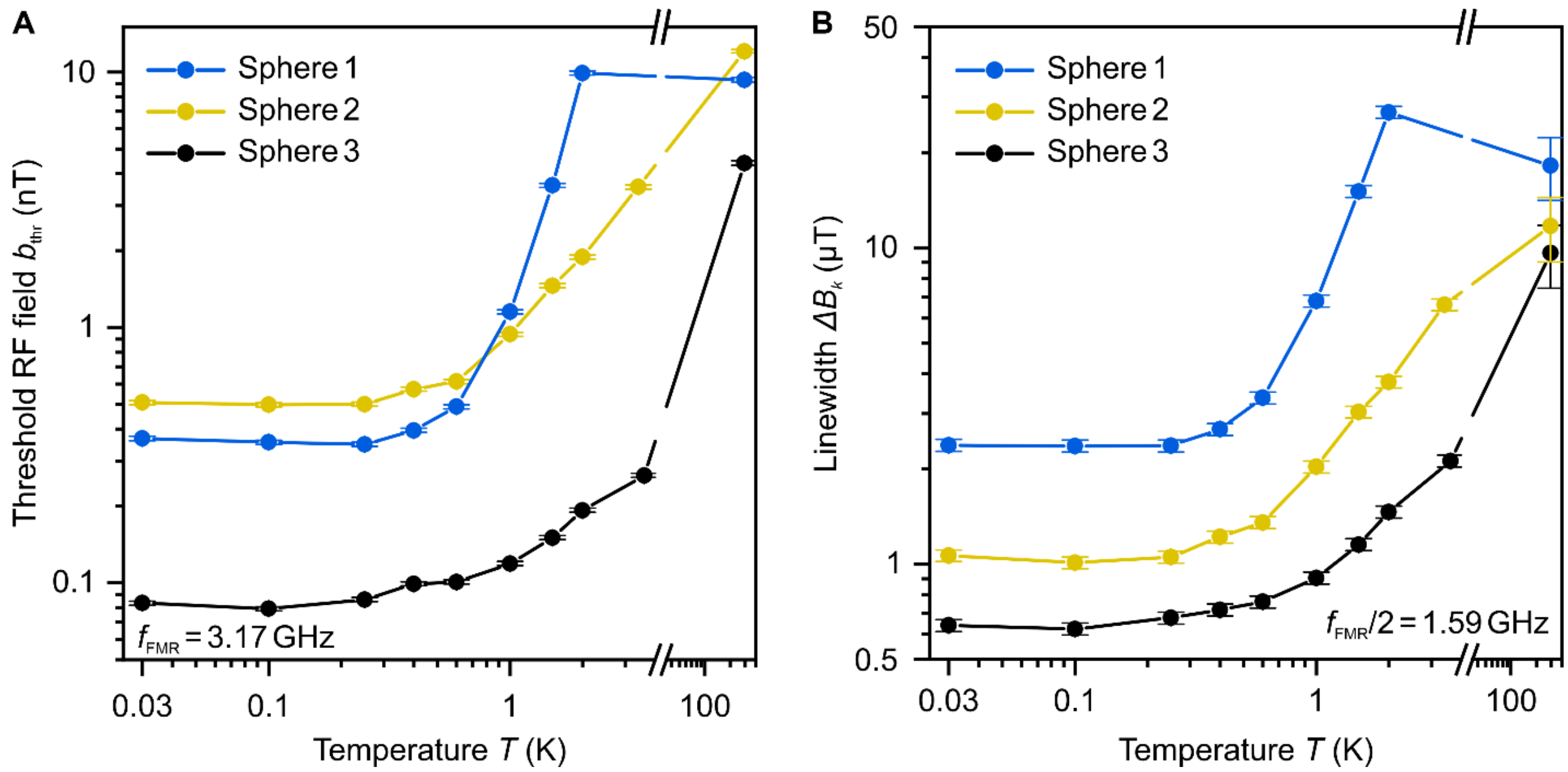

**Fig. S2.**

**Threshold RF field $b_{thr}$ and linewidth $\Delta B_k$ vs temperature $T$.** **A** Threshold RF field $b_{thr}$ as a function of temperature on a logarithmic x-axis for three different YIG spheres at the FMR frequency of 3.17 GHz. **B** Linewidth of secondary magnons $\Delta B_k$ with the frequency half of the FMR frequency vs the temperature $T$ on a logarithmic x-axis for three different YIG spheres. These figures complement Fig. 3 from the main manuscript.

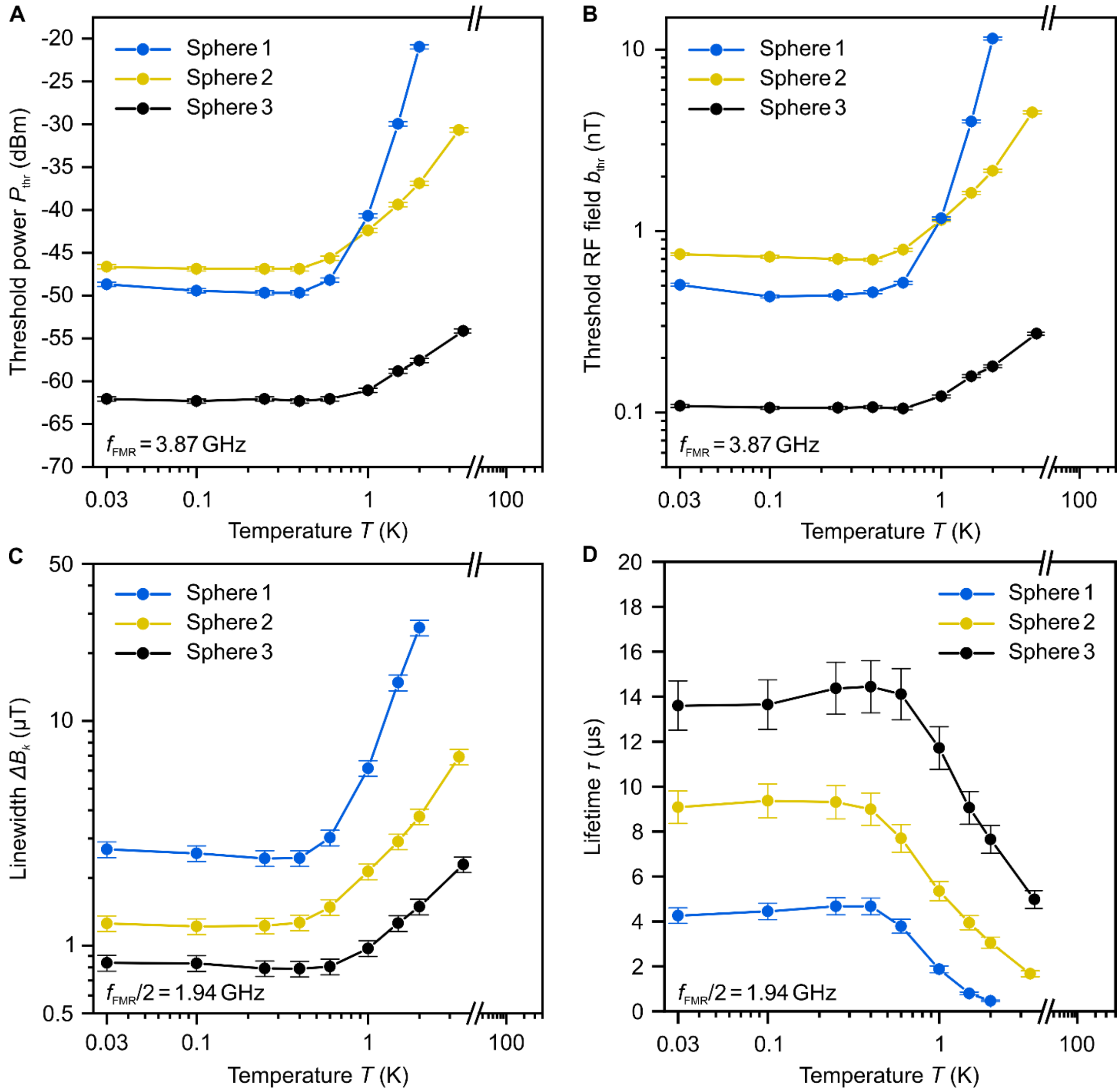

**Fig. S3.**

**Measurement of secondary magnons for $f_{FMR}$ = 3.87 GHz vs temperature. A** Experimental measurements of the parametric power threshold $P_{thr}$ of three different YIG spheres as a function of the temperature $T$ plotted in a logarithmic x-axis for a FMR frequency of 3.87 GHz. **B** Threshold RF field $b_{thr}$ as a function of temperature on a logarithmic x-axis for three different YIG spheres at the FMR frequency of 3.87 GHz. **C** Linewidth of secondary magnons $\Delta B_k$ with the frequency half of the FMR frequency vs the temperature $T$ on a logarithmic x-axis for three different YIG spheres. **D** Lifetime τ of secondary magnons vs the temperature $T$ on a logarithmic x-axis for three different YIG spheres. It was not possible to take measurements at 3.87 GHz at RT because the lower saturation magnetization $M_s$ of YIG at higher temperatures renders the three-magnon scattering process inaccessible at this frequency.

**Table S1.**

**Gilbert fit parameters.** Gilbert damping α and inhomogeneous linewidth broadening $\Delta B_0$ for three different YIG Spheres and two distinct temperatures, 300 K and 30 mK from Fig. S1.

| **Sample:** | **Gilbert damping α ($10^{-5}$)** | | **Inhomogeneous linewidth broadening $\Delta B_0$ (µT)** | |
|---|---|---|---|---|
| **Temperature** | **@ 300 K** | **@ 30 mK** | **@ 300 K** | **@ 30 mK** |
| **Sphere 1** | **8.78 ± 0.26** | **3.92 ± 0.67** | **4.45 ± 2.23** | **8.08 ± 6.01** |
| **Sphere 2** | **3.08 ± 0.50** | **1.01 ± 1.34** | **50.72 ± 4.54** | **59.46 ± 12.47** |
| **Sphere 3** | **6.21± 0.44** | **2.05 ± 0.54** | **-4.35 ± 3.86** | **7.61 ± 4.85** |

**Supplementary References**